# Variational principle for fractional kinetics and the Lévy Ansatz

Sumiyoshi Abe

*Department of Physical Engineering, Mie University, Mie 514-8507, Japan*

**Abstract**

A variational principle is developed for fractional kinetics based on the auxiliary-field formalism. It is applied to the Fokker-Planck equation with spatio-temporal fractionality, and a variational solution is obtained with the help of the Lévy *Ansatz*. It is shown how the whole range from subdiffusion to superdiffusion is realized by the variational solution, as a competing effect between the long waiting time and the long jump. The motion of the center of the probability distribution is also analyzed in the case of a periodic drift.

Let $l$ be the spatial extension of a time-dependent distribution defined by its half width or the root-mean square displacement *if any*. Then, diffusion is characterized by the property that $l$ grows in time as

$$l \sim t^\mu, \qquad (1)$$

where $\mu$ is a positive constant. Normal diffusion observed in ordinary Brownian motion has the value $\mu = 1/2$. Otherwise, it is called anomalous diffusion: subdiffusion if $\mu < 1/2$ and superdiffusion if $\mu > 1/2$.

Anomalous diffusion and related exotic scaling phenomena are of wide interest for various disciplines from the natural to social sciences, and there are a large amount of works devoted to this subject. A very short list of examples includes: particle dispersion in the turbulent atmosphere [1], transport in amorphous solids [2] and disordered media [3], motion of chains of proteins [4], microspheres moving in a living cell [5], mRNA motion [6], endosome trajectories [7], heterogeneous populations of organisms [8], epidemics [9], predator's search behavior [10], earthquakes [11], human travel or mobility patterns [12,13], and trading activity [14].

Fractional kinetics [15,16] is known to offer a unified theoretical framework for modeling anomalous diffusion in terms of widely applicable stochastic processes termed continuous-time random walks. However, it is generically a nontrivial problem to obtain an analytical form of the solution of a given fractional kinetic equation [17-19]. Accordingly, researchers perform numerical analysis (see Ref. [20], for example).

The variational principle, which derives a basic physical equation as the stationarity condition on an action functional, also provides a powerful method for obtaining a nonperturbative analytical solution of the equation. The Rayleigh-Ritz procedure using an *Ansatz* for the form of a variational solution plays a vital role, not limited to quantum mechanics. In particular, its generalization to time-dependent theory can drastically reduce the number of dynamical degrees of freedom contained in a system and thus makes a wide class of problems analytically tractable.

In this paper, we develop a variational principle for kinetic theory containing spatio-temporal fractionality. We employ the auxiliary-field formalism and propose the Lévy *Ansatz* for a probability distribution. This method determines dynamical evolution of the variables defining the *Ansatz* as well as the auxiliary field in terms of the fractional mechanical equations. A periodic drift is considered as an example of interest. We show how the analytic solution exhibits both subdiffusion and superdiffusion. We also discuss the motion of the center of the probability distribution.

Let $p(x,t)$ be a normalized probability distribution defined on $(-\infty, \infty) \times [0, T]$. The fractional Fokker-Planck equation we consider here is given by

$$\frac{\partial p(x,t)}{\partial t} = {}_0D_t^{1-\alpha}\left[-\frac{\partial}{\partial x}\big(F(x)\, p(x,t)\big) - D^*\,(-\Delta)^{\gamma/2} p(x,t)\right]. \qquad (2)$$

$D^*$ is a generalized diffusion coefficient. ${}_0D_t^{1-\alpha}$ and $-(-\Delta)^{\gamma/2}$ are the Riemann-Liouville fractional differential operator and the Riesz fractional "Laplacian" operator, respectively (see Refs. [21,22], where slightly different notations are used). We assume that the indices satisfy the following conditions:

$$0 < \alpha < 1, \qquad 0 < \gamma < 2. \tag{3}$$

Then, the Riemann-Liouville operators are defined as follows: ${}_0D_t^{1-\alpha} f(t) = [1/\Gamma(\alpha)]$ $\times (d/dt) \int_0^t d\tau \, (t-\tau)^{\alpha-1} f(\tau)$, ${}_tD_T^{1-\alpha} f(t) = [1/\Gamma(\alpha)](-d/dt) \int_t^T d\tau \, (\tau-t)^{\alpha-1} f(\tau)$, where $\Gamma(\alpha)$ is the Euler gamma function. They satisfy the relation: $\int_0^T dt \left[ {}_0D_t^{1-\alpha} f(t) \right] g(t)$ $= \int_0^T dt \, f(t) \left[ {}_tD_T^{1-\alpha} g(t) \right]$. On the other hand, the Riesz operator is given by $-(-\Delta)^{\gamma/2}$ $\equiv -[2\cos(\pi\gamma/2)]^{-1} \left[ d^\gamma/dx^\gamma + d^\gamma/d(-x)^\gamma \right]$, which satisfies $-(-\Delta)^{\gamma/2} \exp(\pm ikx)$ $= -|k|^\gamma \exp(\pm ikx)$ and $\int_{-\infty}^\infty dx \left[ -(-\Delta)^{\gamma/2} \psi(x) \right] \phi(x) = \int_{-\infty}^\infty dx \, \psi(x) \left[ -(-\Delta)^{\gamma/2} \phi(x) \right]$.

The derivative appearing in the drift term could also be fractionalized, in general (see Ref. [15]). In the random walk picture [23], the temporal and spatial fractionalities in Eq. (2) imply [16,24] that both the waiting-time and jump probabilities obey the power laws characterized by $\alpha$ and $\gamma$, respectively. More precisely, the conditions in Eq. (3) imply that both the average waiting time and the second moment of the jump are divergent.

Let us proceed to formulate a variational principle for Eq. (2). Since the left-hand side of the equation is the first-order derivative in time, it is convenient to employ the auxiliary-field formalism [25,26]. The action reads

$$I = -\int_0^T dt \left\langle \dot{\Lambda} + {}_tD_T^{1-\alpha} \left[ F \Lambda' - D^* (-\Delta)^{\gamma/2} \Lambda \right] \right\rangle - \left\langle \Lambda \right\rangle \big|_{t=0}, \tag{4}$$

where $\Lambda = \Lambda(x,t)$ is an auxiliary field, $\langle A \rangle = \int_{-\infty}^\infty dx \, A(x,t) \, p(x,t)$, $\dot{A} = \partial A/\partial t$, and $A' = \partial A / \partial x$. It is noted that the auxiliary field is not a probability distribution.

The variations of this action with respect to $\Lambda$ and $p$ lead to

$$\delta_\Lambda I = \int_0^T dt \int_{-\infty}^\infty dx \left\{ \dot{p} + {}_0D_t^{1-\alpha}\left[(Fp)' + D^*(-\Delta)^{\gamma/2} p\right] \right\} \delta\Lambda - \int_{-\infty}^\infty dx\, p\, \delta\Lambda \bigg|_{t=T}, \quad (5)$$

$$\delta_p I = -\int_0^T dt \int_{-\infty}^\infty dx \left\{ \dot{\Lambda} + {}_tD_T^{1-\alpha}\left[F\Lambda' - D^*(-\Delta)^{\gamma/2} \Lambda\right] \right\} \delta p - \int_{-\infty}^\infty dx\, \Lambda\, \delta p \bigg|_{t=0}, \quad (6)$$

respectively. The temporal boundary terms appearing in these equations can be eliminated if $p(x,0)$ and $\Lambda(x,T)$ are taken to be fixed functions of $x$. Then, from Eqs. (5) and (6), follow Eq. (2) and

$$\frac{\partial \Lambda(x,t)}{\partial t} = - {}_tD_T^{1-\alpha}\left[ F(x)\frac{\partial \Lambda(x,t)}{\partial x} - D^*(-\Delta)^{\gamma/2} \Lambda(x,t) \right], \quad (7)$$

respectively.

Now, as a drift term, we consider in the present work a periodic one as an example of interest [20,27,28]. In particular, we take the following form:

$$F(x) = F_0 \sin(Kx), \quad (8)$$

where both $F_0$ and $K$ are constants and are taken to be positive without loss of generality.

The temporal boundary conditions we impose on the probability distribution and the auxiliary field are given as follows:

$$p(x, 0) = \delta(x - X_0), \tag{9}$$

$$\Lambda(x, T) = \cos(\kappa(x - \xi_T)), \tag{10}$$

where $X_0$ and $\xi_T$ are constants, and $\kappa$ is a *small* parameter.

To obtain a variational solution of Eq. (2), first we propose the Lévy *Ansatz* for the probability distribution:

$$p(x,t) \equiv p(x,t; X(t)) = \frac{1}{2\pi} \int_{-\infty}^{\infty} dk \exp[-ik(x - X(t))] \exp(-a(t)|k|^\gamma), \tag{11}$$

where $a(t)$ is a nonnegative function of time. This *Ansatz* is motivated by the property of the Riesz operator appearing in Eq. (2). It satisfies the following scaling law:

$$p(x,t; X(t)) = \frac{1}{a^{1/\gamma}(t)} \tilde{p}\left(\frac{x}{a^{1/\gamma}(t)}; \frac{X(t)}{a^{1/\gamma}(t)}\right), \tag{12}$$

where $\tilde{p}(x; X) \equiv (1/2\pi) \int_{-\infty}^{\infty} ds \exp[-is(x - X) - |s|^\gamma]$ is a scaling function. Then, the spatial extension such as the half width, $l$, is seen to scale as

$$l \sim a^{1/\gamma}(t). \tag{13}$$

Eq. (11) is a Lévy distribution centered at $x = X(t)$, which decays as a power law,

$p(x,t) \sim |x|^{-\gamma-1}$, for large $|x|$. Since its second moment is divergent, it is not allowed to choose as the auxiliary field a polynomial of the position variable. This is in contrast to the Gaussian *Ansatz* [26] discussed in ordinary kinetics, i.e., the nonfractional one. An allowed form for $\Lambda(x,t)$ also depends on the property of a given drift term. (We will mention this point again in the concluding part of the paper.) In the present case of the periodic drift in Eq. (8), the following form turns out to be useful:

$$\Lambda(x,t) = \lambda(t)\cos\left[\kappa\left(x - \xi(t)\right)\right]. \tag{14}$$

Thus, our Rayleigh-Ritz-like variational method contains four variables: $a(t)$, $X(t)$, $\lambda(t)$, and $\xi(t)$. From the temporal boundary conditions in Eqs. (9) and (10), these variables have to satisfy

$$a(0) = 0, \qquad X(0) = X_0, \tag{15}$$

$$\lambda(T) = 1, \qquad \xi(T) = \xi_T. \tag{16}$$

Substitution of Eqs. (11) and (14) into Eq. (4) yields, up to irrelevant temporal boundary terms, the reduced action, $I = \int_0^T dt\, L$, where $L$ is the Lagrangian given by

$$\begin{aligned}
L = &-\lambda(t)\left\{\kappa \dot{X}(t)\sin\left[\kappa\left(X(t)-\xi(t)\right)\right] + |\kappa|^\gamma \dot{a}(t)\cos\left[\kappa\left(X(t)-\xi(t)\right)\right]\right\} \\
&\times \exp\left(-a(t)|\kappa|^\gamma\right) \\
&+ \frac{1}{4}\kappa F_0 \lambda(t)\left\{\exp(-i\kappa\xi(t))\,_0D_t^{1-\alpha}\left[\exp\left(i(K+\kappa)X(t)-a(t)|K+\kappa|^\gamma\right)\right]\right. \\
&\left. -\exp(i\kappa\xi(t))\,_0D_t^{1-\alpha}\left[\exp\left(i(K-\kappa)X(t)-a(t)|K-\kappa|^\gamma\right)\right]\right\}
\end{aligned}$$

$$-\exp(-i\kappa\xi(t))\,_0D_t^{1-\alpha}\left[\exp\left(-i(K-\kappa)X(t)-a(t)|K-\kappa|^\gamma\right)\right]$$

$$+\exp(i\kappa\xi(t))\,_0D_t^{1-\alpha}\left[\exp\left(-i(K+\kappa)X(t)-a(t)|K+\kappa|^\gamma\right)\right]\}$$

$$+\frac{1}{2}|\kappa|^\gamma D^*\lambda(t)\left\{\exp(-i\kappa\xi(t))\,_0D_t^{1-\alpha}\left[\exp\left(i\kappa X(t)-a(t)|\kappa|^\gamma\right)\right]\right.$$

$$\left.+\exp(i\kappa\xi(t))\,_0D_t^{1-\alpha}\left[\exp\left(-i\kappa X(t)-a(t)|\kappa|^\gamma\right)\right]\right\}. \quad (17)$$

The variables, $a(t)$, $X(t)$, $\lambda(t)$, and $\xi(t)$, are now fixed at the boundaries, $t=0$ and $t=T$. Recalling that $\kappa$ is a small parameter, we obtain in the leading orders the following Euler-Lagrange equations for these variables:

$$\dot\lambda(t)=0, \quad (18)$$

$$\dot\lambda(t)(X(t)-\xi(t))-\lambda(t)\dot\xi(t)$$
$$=F_0\left\{\left(1-\gamma a(t)|K|^\gamma\right)\sin(KX(t))+KX(t)\cos(KX(t))\right\}$$
$$\times\exp\left(-a(t)|K|^\gamma\right)\,_tD_T^{1-\alpha}[\lambda(t)]$$
$$-F_0 K\cos(KX(t))\exp\left(-a(t)|K|^\gamma\right)\,_tD_T^{1-\alpha}[\lambda(t)\xi(t)], \quad (19)$$

$$\dot a(t)=D^*\,_0D_t^{1-\alpha}[1], \quad (20)$$

$$\dot X(t)+F_0\,_0D_t^{1-\alpha}\left[\sin(KX(t))\exp\left(-a(t)|K|^\gamma\right)\right]=0. \quad (21)$$

Equations (18) and (20) can immediately be solved. The solution of Eq. (18) satisfying the condition in Eq. (16) is

$$\lambda(t)=1. \quad (22)$$

Therefore, in the limit $\kappa\to 0$, we have $\Lambda(x,t)=1$, which automatically leads to the subsidiary condition [26]: $\langle \Lambda \rangle = 1$. The solution of Eq. (20) satisfying the condition in Eq. (15) is found to be

$$a(t) = \frac{D^*}{\Gamma(\alpha+1)} t^\alpha. \qquad (23)$$

On the other hand, Eqs. (19) and (21) are involved. To see the property of the probability distribution, it is sufficient to analyze only Eq. (21), here.

Equation (21) is invariant under the shift: $X(t) \to X(t) + 2\pi n/K$, where $n$ is an arbitrary integer. This symmetry correctly reflects the periodicity of the drift in Eq. (8). Accordingly, the probability distribution is also periodic: $p(x + 2\pi/K, t) = p(x, t)$. To evaluate the behavior of the solution, we take the initial value, $X_0$, in the vicinity of the origin and consider the "long wavelength limit", in which $K$ is small. In such a limit, Eq. (21) becomes

$$X(t) + \frac{KF_0}{\Gamma(\alpha)} \int_0^t d\tau \, (t-\tau)^{\alpha-1} X(\tau) = X_0. \qquad (24)$$

This equation can be solved by the use of the Laplace transformation. The solution is

$$X(t) = X_0 \, E_\alpha\left(-KF_0 t^\alpha\right), \qquad (25)$$

where $E_\alpha(z)$ is the Mittag-Leffler function [21,22] defined by $E_\alpha(z) = \sum_{n=0}^\infty z^n / \Gamma(\alpha n + 1)$. Therefore, we see that the center of the probability distribution approaches the origin as $X(t) \sim t^{-\alpha}$, which is very slow compared to the ordinary exponential case corresponding to the limit $\alpha \to 1$ [note that $E_1(z) = \exp(z)$].

Thus, we have specified the variational solution of the fractional Fokker-Planck equation (2) with the periodic drift in Eq. (8). From Eqs. (13) and (23), consequently we have

$$l \sim t^{\alpha/\gamma}, \qquad (27)$$

which implies that the whole range from subdiffusion to superdiffusion is covered. In particular, normal diffusion is realized if $\gamma = 2\alpha$, which is consistent with the conditions in Eq. (3). This feature has a simple interpretation. Diffusion is suppressed by a long waiting time (temporal fractionality) and is enhanced by a long jump (spatial fractionality). The covering of both subdiffusion and superdiffusion results from such competing effects.

In conclusion, we have studied the variational principle for fractional kinetics based on the auxiliary field formalism. We have developed the Rayleigh-Ritz-like method for the Fokker-Planck equation with spatio-temporal fractionality as well as a periodic drift and obtained the analytical form of a variational solution based on the Lévy *Ansatz*. We have seen how the whole range from subdiffusion to superdiffusion can continuously be realized in a unified way by the variational solution. In addition, we have analyzed the motion of the center of the probability distribution. Besides a form of the probability distribution, an *Ansatz* for the auxiliary field also plays a crucial role. In the present model with the periodic drift, we have employed the auxiliary field that is also periodic.

An allowed form for the auxiliary field, however, depends on the property of a drift term. For example, if the Lévy *Ansatz* is employed in a model with a drift consisting of a positive power of the position variable, then the auxiliary field should decay fast enough at spatial infinities in order for the action to be finite.

This work has been supported in part by a Grant-in-Aid for Scientific Research from the Japan Society for the Promotion of Science.

______________________

(E) **47**, 1765 (1981).

[26] O. Éboli, R. Jackiw, and S.-Y. Pi, Phys. Rev. D **37**, 3557 (1988).

[27] H. Risken, *The Fokker-Planck Equation, 2nd Edition*

(Springer-Verlag, Berlin, 1989).

[28] P. Hänggi, F. Marchesoni, adn F. Nori, Ann. Phys. (Leipzig) **14**, 51 (2005).13